\documentclass[aps,pre,twocolumn,showpacs,floatfix,preprintnumbers,amsmath,amssymb]{revtex4-1}
\usepackage{amsmath}
\usepackage{color}
\usepackage{amssymb}
\usepackage{graphicx}
\usepackage{epsfig}
\usepackage{dcolumn}
\usepackage{bm}
\usepackage{array}
\usepackage[caption=false]{subfig}

\begin{document}

\title{Distinct nodes visited by random walkers on scale-free networks}
\author{Aanjaneya Kumar and M. S. Santhanam}
\affiliation{Indian Institute of Science Education and Research, \\
Dr. Homi Bhabha Road, Pune 411008, India.}

\begin{abstract}
Random walks on discrete lattices are fundamental models that form the basis 
for our understanding of transport and diffusion processes. For a single random 
walker on complex networks, many properties such as the mean first passage 
time and cover time are known. However, many recent applications such as search 
engines and recommender systems involve multiple random walkers on complex networks.
In this work, based on numerical simulations, we show that the fraction of nodes 
of scale-free network not visited by $W$ random walkers in time $t$ has a 
stretched exponential form independent of the details of the network and number
of walkers. This leads to a power-law relation between nodes not visited by $W$ 
walkers and by one walker within time $t$. The problem of finding the distinct nodes visited by 
$W$ walkers, effectively, can be reduced to that of a single walker. The robustness
of the results is demonstrated by verifying them on four different real-world
networks that approximately display scale-free structure.

\end{abstract}


\maketitle

Random walks were introduced more than a century ago and have formed the basis 
for our understanding of diffusion processes in physical systems \citep{rw}. As a
fundamental stochastic process, they are relevant for many fields ranging 
from physics and computer sciences \citep{web} to biology \citep{rwbio} 
and economics \citep{eco}. Several
problems including animal foraging and migration \citep{forag,migr},
emergence of innovation \cite{inno}, 
intracellular molecular transport \cite{insulin},
proteins binding with DNA sequences \cite{dna1}, for structual
information about macromolecules \cite{bio} are based on the dynamics of a single
random walker on regular lattice or its variants.

On the other hand, many recent applications 
involve dynamics of {\sl multiple} random walkers on a disordered lattice, e.g., complex
network with non-local edges connecting the nodes.
For instance, cellular signal transduction \citep{cell}, exciton transport in molecular
crystals, web search algorithms \citep{web}, a class of image segmentation 
algorithms \cite{imseg}, graph clustering \cite{gclust} and 
recommender systems \citep{llu} widely used for personalization in popular websites are based on 
the idea of many random walkers exploring a topology of discrete nodes connected
through their edges. 

As a statistical physics problem, in comparison to the well-studied 
problem of the dynamics of a {\sl single} random walker on regular lattice \cite{mont} 
or complex network \cite{porter, burioni}, the case of multiple walkers in a network 
setting has not attracted sufficient research attention.
In random walk with $W$ non-interacting walkers, some results are a straightforward
generalisation of that for single walker dynamics. For instance, on a complex network, occupation
probability of a single walker on a node with degree $k$ is proportional to $k$, whereas
for $W$ walkers it is $\propto W k$. However, in many cases, the results for multiple 
walker dynamics is not a trivial generalization of that for a single walker.
One such statistical quantity of interest is the mean number of distinct
sites $S_W(t)$ visited by $W$ random walkers in $t$ discrete time steps on a network with
$N$ nodes. This is relevant for problems related to (mis-)information and contagion spreading
and search problems on networks \citep{epi}.

Distinct sites visited in $t$-steps by a random walker was studied in Refs. \citep{de} 
and its generalization to $W$ walkers was considered in Refs. \citep{stan,weiss}. 
On regular $d$-dimensional
lattices and for short times, the mean number of distinct sites visited by $W$ walkers is
$\langle S_W(t) \rangle \propto t^{d}$ and asymptotically $\langle S_W(t) \rangle \propto Wt$ 
for $d>3$. In general, Ref. \citep{stan} identifies three distinct time scales
with different behaviours for $\langle S_W(t) \rangle$ and limited analytical support
is presented in Ref. \citep{yuste}. Recently, an exact asymptotic
result for the distribution of number of distinct and common sites visited by $W$ walkers 
on a regular 1$d$ lattice was obtained \citep{snm1} by transforming it as a problem of 
extreme value statistics.

In spite of these developments for regular lattices, very few results are known for multiple random
walkers on complex networks. For a random walker on a Bethe lattice with 
coordination number $z$, $ S_1(t) = ((z-2)/(z-1))t$, for $z \ge 3$ \citep{bethe}.
Clearly, $S_1(t) \propto n$ with a prefactor that depends on the local topology
of the lattice.  On a random network, a formal relation for the
generating function corresponding to $S_1(t)$ has been obtained in terms
of the generating functions for the first passage probabilities \citep{snm2}. For a walker
on scale-free network, it was numerically shown that, for short times, $S_1(t)=t$
and as $t\to\infty, S_1(t) \to 1$ due to finite size of network \citep{slee}.
On a small world network, $S_1(t)$ displays a cross-over from $\sqrt{t}$
to linear behaviour depending on whether the walker has managed
to hit a short-cut in the small world network or not \citep{kulk}.

To the best of our knowledge, exact closed form result for $\langle S_W(t) \rangle$
on an arbitrary network with $N$ nodes is not yet known. Even as this gap continues
to exist, in this work, new results primarily based on numerical simulations are presented
that effectively relate $S_W(t)$ to $S_1(t)$ on static networks.
In particular, for the class of scale-free networks, it is shown that the number
of nodes $s_W(t)$ not yet visited until time $t$ has a stretched exponential form, with
exponent $\beta$, depending on the specific network structure but independent of
the number of walkers.
As shown below, the results are consistent with $s_W(t) \propto \exp( W^\beta s_1(t) )$.
Thus, effectively, the problem of finding $S_W(t)$ can be 
reduced to a relatively simpler problem of finding $S_1(t)$ on a scale-free network.

\begin{figure}[t]
\subfloat[]{\includegraphics*[width=3in]{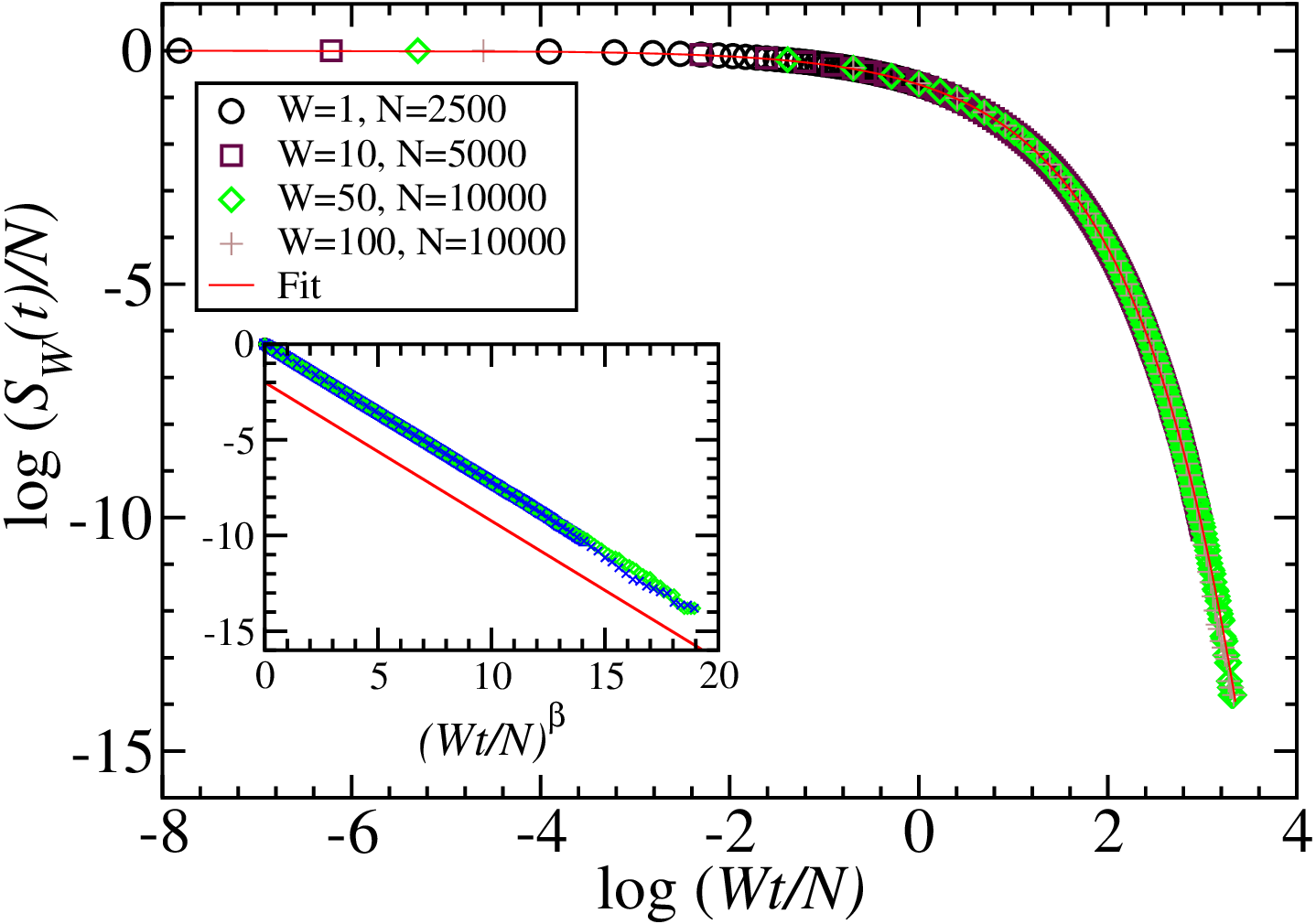}}

\subfloat[]{\includegraphics*[width=3.2in]{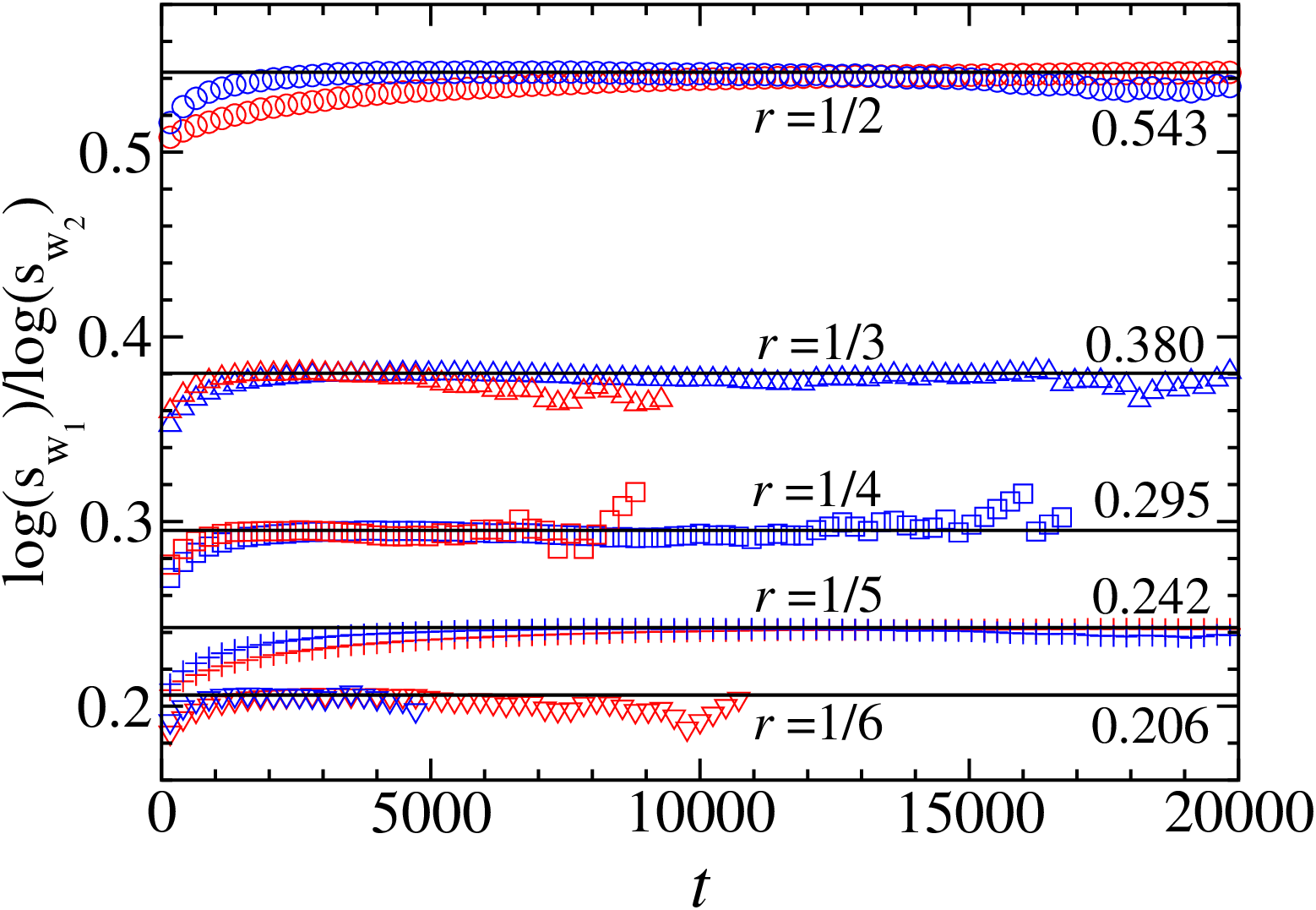}}
\caption{(a) Fraction of nodes not reachable by random walkers on Barabasi-Albert
scale-free network plotted as a function of $x=Wt/N$. Symbols are from
random walk simulations and were averaged over 1000 realizations.
The solid (red) curve is the best fit line. Note the excellent scaling collapse.
(Inset) shows the same data (in semi-log scale) as a function of $(Wt/N)^{\beta}$.
The best fit (solid) line with slope $\beta=0.693$ is given a vertical offset
for easier comparison with simulation data. (b) The ratio $u(t)$ shown as a function
of $t$. Validity of scaling can be inferred from the flat lines obtained from simulations.
In this, $r=W_1/W_2$ and the value of $r^\beta$ is also indicated.}
\label{unr1}
\end{figure}

Distinct sites visited by multiple walkers $S_W(t)$ can also be thought of as 
a statistical relaxation process especially if the initial position 
of the walkers is far from equilibrium distribution. On a network, this is
easily achieved by placing all the walkers on the same node at $t=0$.
In this garb, $s_W(t) = N(1-S_W(t)/N)$ represents a relaxation process and the
results presented here indicate scaling of this process as a function of $W$ and $N$.
Such relaxation processes in a wide variety of disordered condensed matter systems
is known to display a stretched exponential decay of auto-correlations of the form
$C(t) = \exp(-t/\tau)^\beta$, where $\tau$ is a parameter with dimensions of 
inverse time and $0 \le \beta \le 1$ is the exponent \citep{klafter}. The results obtained
in this work add to the list of known systems that display stretched exponential relaxation.
If $S_W(t)$ denotes the unique 'territory' covered by $W$ walkers, then $s_W(t)$ represents
its complementary part, the territory unreachable in time $t$. Thus $s_W(t) + S_W(t)=1$
for all $t$ and we have chosen to present results for $s_W(t)$ in the rest of the paper.

\begin{figure}[t]
\includegraphics*[width=3.2in]{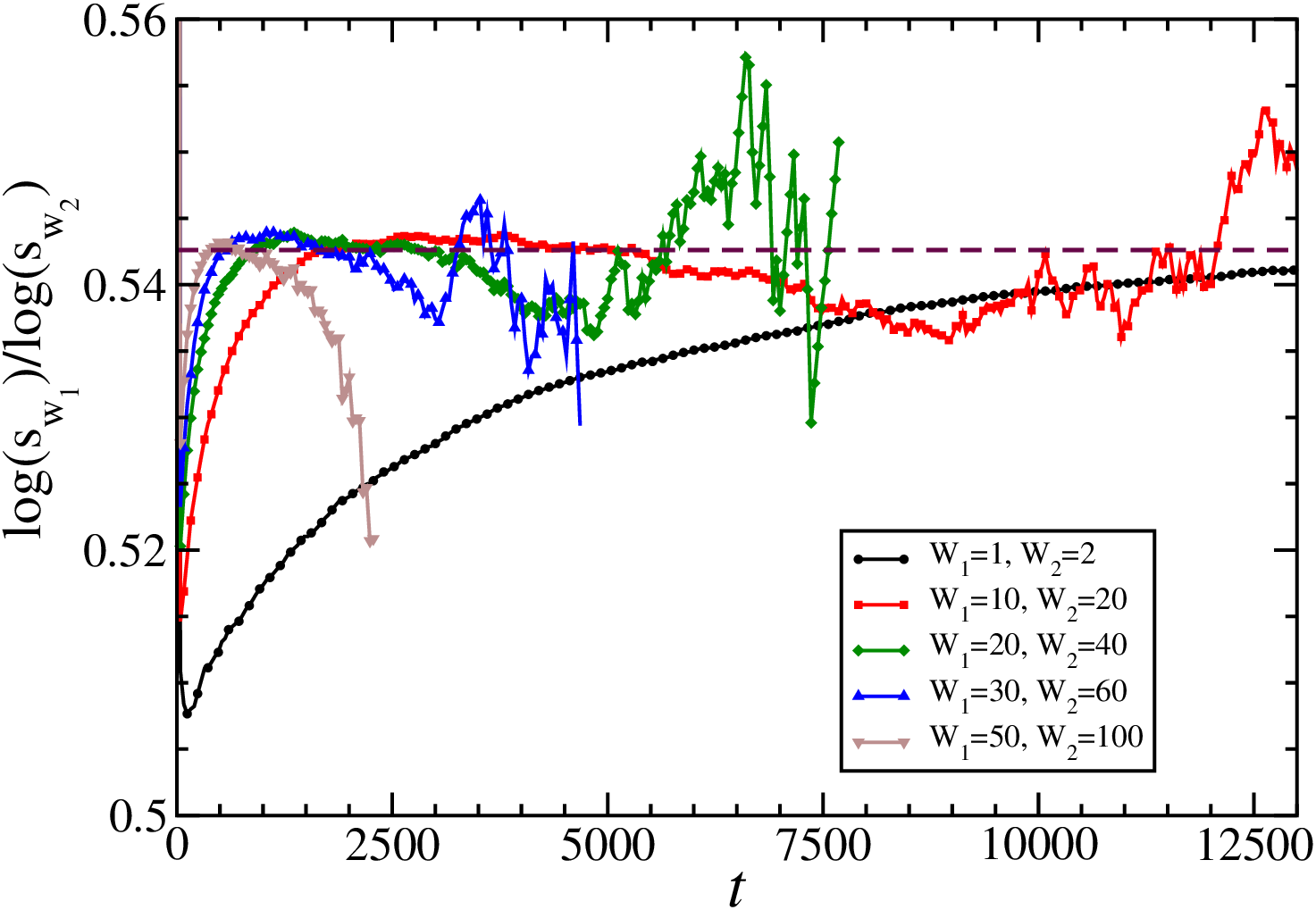}
\caption{The ratio $u(t)$ shown as a function of $t$ for several choices
of $W_1$ and $W_2$. The deviations from the scaling relation in Eq. \ref{scaling}
arise due to finite number of walkers and finite size of the network.
See text for a detailed explanation. This simulations were performed on Barabasi-Albert
scale-free network with $N=10000$ nodes (same as one of the networks used in Fig. \ref{unr1}(a)).}
\label{dev1}
\end{figure}

Random walks on complex networks are a straightforward generalisation
of random walks on regular lattices. In this work, independent and multiple
walkers randomly walk on a connected, scale-free network with $N$ nodes and $E$ edges
generated using Barabasi-Albert (BA) \citep{scale} and configuration models \citep{newman}. Each node has an associated
degree $k_i, i=1,2...N$, indicating the number of edges. The degree distribution
of the network is $P(k) \sim k^{-\gamma}$, where $\gamma$ is the exponent.
A walker at $i$-th node can hop to any
of its connected neighbours with probability $1/k_i$. The information on the edges
in the network are encoded in the adjacency matrix ${\mathbf A}$ of order $N$,
where the element $A_{ij}=1$ if nodes $i$ and $j$ are connected by an edge,
and $A_{ij}=0$ if they are not connected.

In Fig. \ref{unr1}(a), the fraction of nodes not reached in time $t$, $s_W(t)$,
is shown for BA scale-free networks ${\mathcal N}_{BA}(N,E,\gamma)$.
Each data point is averaged over 1000 random walk realizations.
At time $t=0$, all the $W$ walkers are all placed on a randomly chosen node
designated as zeroth node, {\it i.e.}, $w_0(t=0) = \sum_i W \delta_{i,0}$.
This figure shows results for four different values of $N$ and $W$. Remarkably, 
in all the cases, the scaled parameter can be identified as $x=Wt/N$.
With this choice, scaling is evident from the excellent data collapse observed 
in Fig. \ref{unr1}(a). It can be inferred (from the fitted solid line) that as $W \to \infty$
and $N \to \infty$,
the fraction of unreachable nodes is consistent with a stretched exponential function 
of the form,
\begin{equation}
s_W(t)= \left( \frac{N-1}{N} \right) ~ e^{-A ~x(t)^{\beta}},
\label{ure}
\end{equation}
in which the parameters $A$ and $\beta$ are estimated through 
a regression procedure. For the simulations shown in Fig. \ref{unr1}, $A \approx 0.75$
and $\beta \approx 0.88$. The scaled time $x=Wt/N$ can  also
be expressed in units of mean relaxation time as $x = C_\beta ~t / \tau_r$, where
$\tau_r = C_\beta N/W$ and $C_\beta=\frac{A ~\Gamma(2/\beta)}{\beta ~\Gamma(1+1/\beta)}$.
For single walker dynamics, $W=1$ and scaled time reduces to $x=t/N$, in agreement
with the results in Ref. \citep{slee}.
The inset in Fig. \ref{unr1}(a) displays the same data as in the main figure
as a function of $x^{\beta}$ and its linearity suggests Eq. \ref{ure}.
For $x \ll 1$, Eq. \ref{ure} becomes $s_W(t) \approx 1-Ax^{\beta}$.
Thus, the fraction of distinct sites visited
is $S_W(t)\approx A x^{\beta}$. 

\begin{figure}[t]
\includegraphics[width=3in]{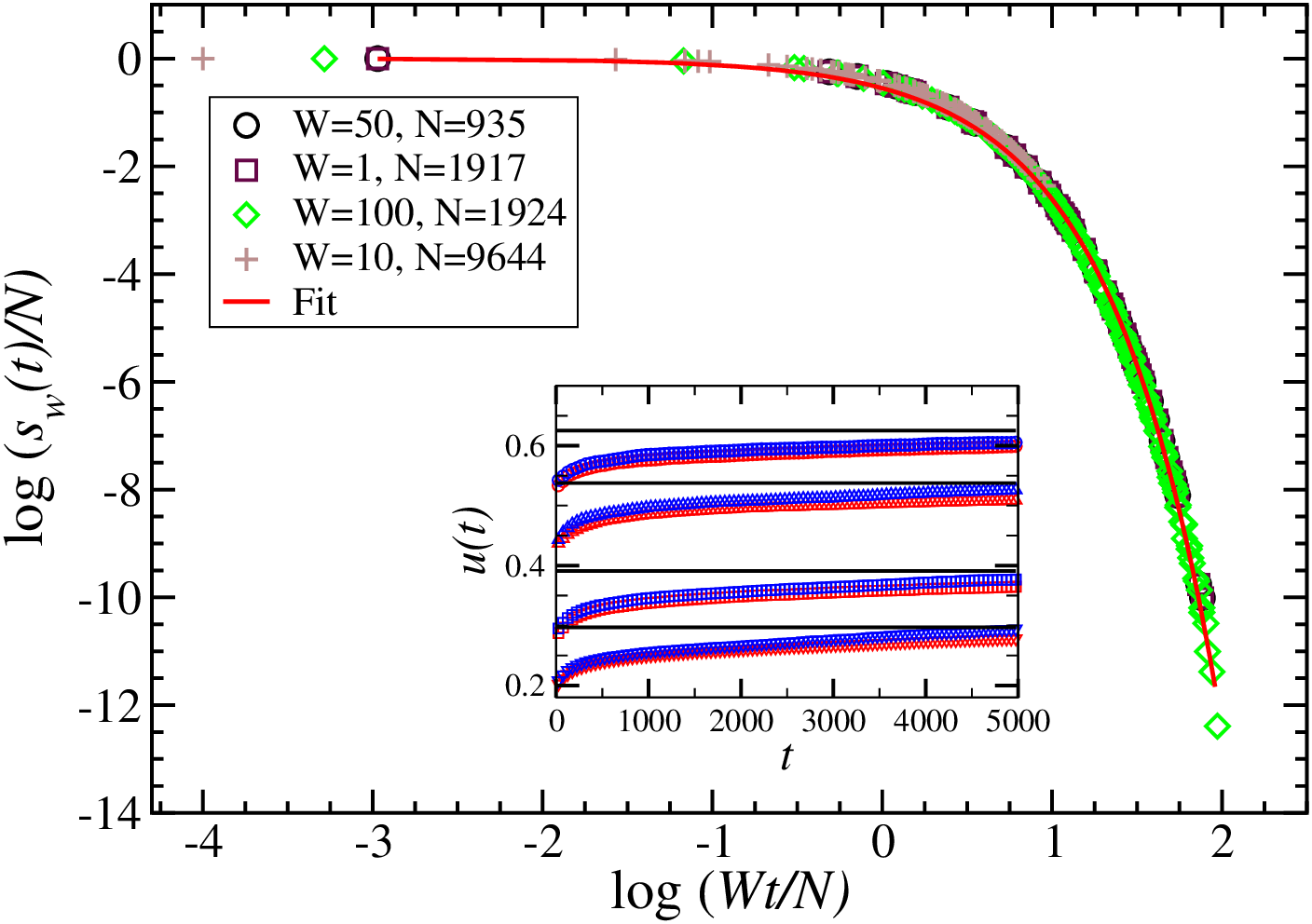}
\caption{Fraction of nodes not reachable by random walkers on configuration
model of scale-free networks plotted as a function of $x=Wt/N$.
Symbols are obtained from random walk simulations averaged over 1000 realizations.
The solid (red) line represents Eq. \ref{ure} with $A=0.601$ and $\beta=0.653$ obtained
through regression. Note the excellent scaling collapse.
(Inset) demonstrates the scaling of $s_w(t)$ (Eq. \ref{scaling}) for different
number of walkers $w_1$ and $w_2$. The solid (horizontal) lines are the expected
value, namely, $(W_2/W_1)^{\beta}$ for $W_2/W_1$ ratios $1/2$, $2/5$, $1/3$ and $1/4$.}
\label{flogw}
\end{figure}

In connected and finite size networks, unreachable 
nodes is a finite time effect since as $t \to \infty$ all the nodes are eventually
reached. Then, we can expect the scaling relation in Eq. \ref{ure} to hold
good in the timescale $\tau_r < t \ll t_{cov}$,
where $t_{cov}$ is the cover time for all the nodes to be visited at least once.
For a single walker on a scale-free network, $t_{cov} \le N \log N$, though a 
similar result for multiple walkers is not yet known \citep{alon}.
Since multiple walkers are known to improve efficiency of covering network \citep{alon},
in this case, $N \log N$ will essentially be a loose upper bound.

Based on Eq. \ref{ure}, the central result of this paper can be recast
in the form of a scaling relation as $N \to \infty$, $W_1,W_2 \to \infty$ 
and it is of the form
\begin{equation}
\log s_{W_2}(t) = \left( \frac{W_2}{W_1} \right)^\beta ~ \log s_{W_1}(t).
\label{scaling}
\end{equation}
This is valid for $W_1, W_2 \gg 0$ random walkers on a given scale-free network
${\mathcal N}_s(N,E,\gamma)$. Remarkably, the relation between 
$s_{W_1}$ and $s_{W_2}$ depends only on the ratio $r=W_1/W_2$ and information 
about network enters through the exponent $\beta$. This is verified in 
Figure \ref{unr1}(b) by plotting the ratio $u(t) = \log s_{W_1}(t)/\log s_{W_1}(t)$ as a 
function of time $t$. In this form, presence of scaling is inferred from horizontal 
lines such that $u(t)=\phi=r^\beta$, a constant.
In particular, simulations confirm that $\phi$ is identical for any
choice of $W_1$ and $W_2$ such that $r$ is a constant.

\begin{figure}[t]
\subfloat[]{\includegraphics*[width=2.5in]{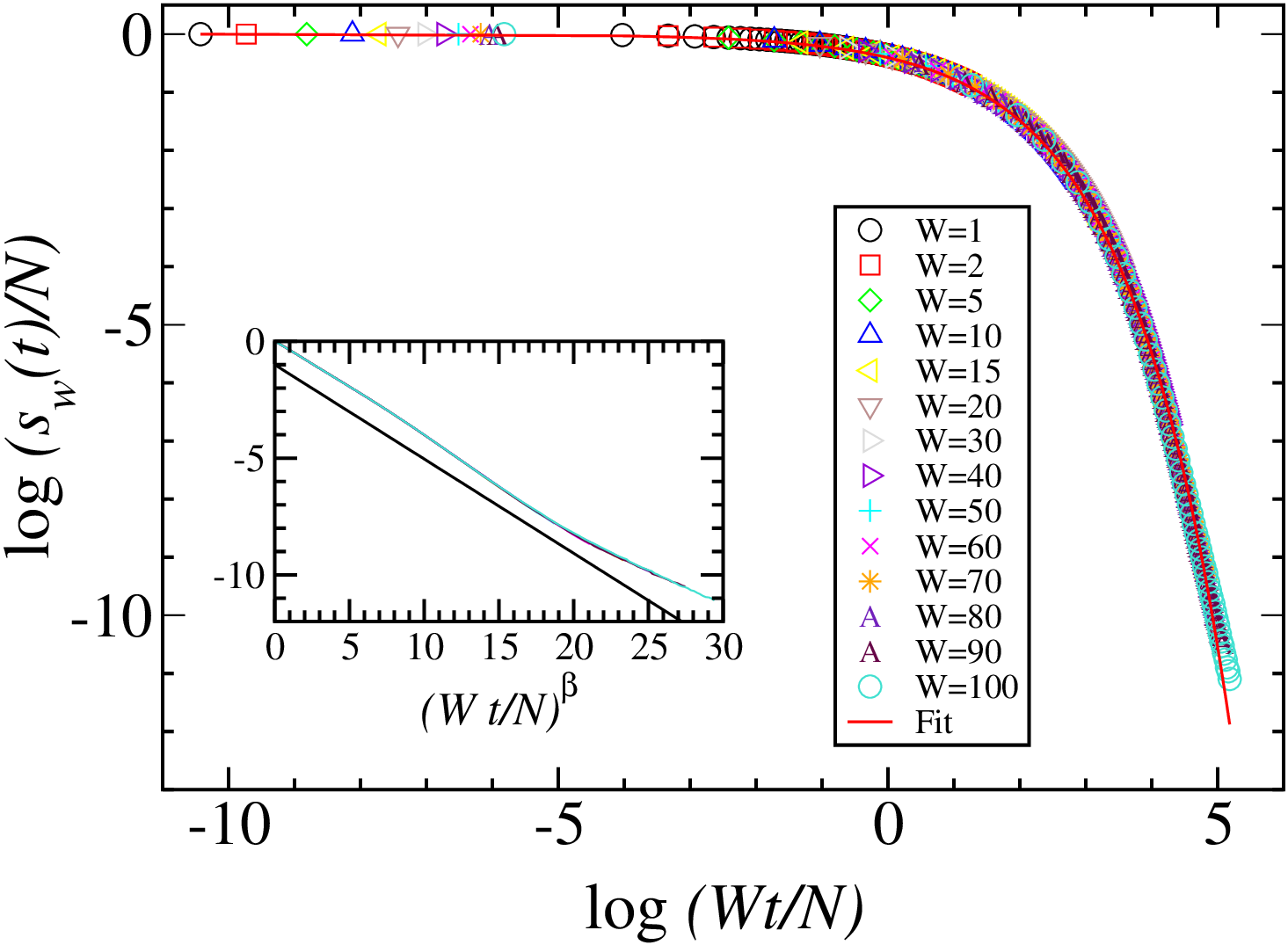}}

\subfloat[]{\includegraphics*[width=2.5in]{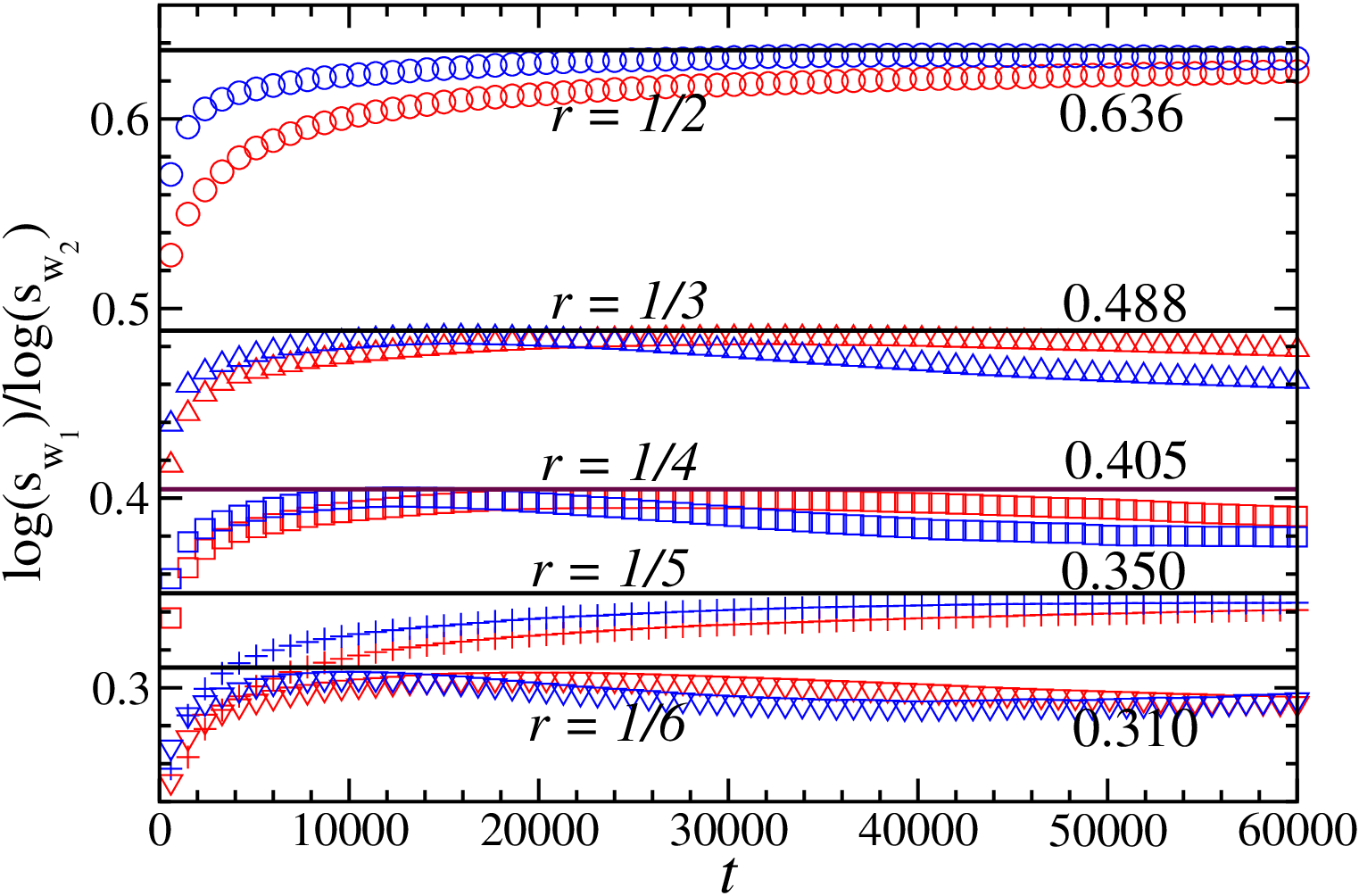}}
\caption{(a) For Enron email network, the fraction of nodes not reached by random walkers $s_W(t)$
as a function of scaled parameter $x$ (in log-log scale) for several values of $W$. The data collapse
points to an agreement with Eq. \ref{ure} with $A=0.403$ and $\beta =0.652$. (Inset) shows
that the same data becomes linear in semi-log plot if plotted as a function of $(Wt/N)^\beta$.
(b) Demonstrates of the scaling relation in Eq. \ref{scaling} for various choices of $r=W_1/W_2$.}
\label{enron-fig}
\end{figure}

The deviations observed in Fig. \ref{unr1}(b) in the vicinity of $t=0$ arise due to the finite number
of walkers on the network. As shown in Fig. \ref{dev1}, as $W \to \infty$
the agreement with the scaling curve in Eq. \ref{scaling} gets better.
On the other hand, the deviations observed in Fig. \ref{unr1}(b) for $t >> 1$
arise due to the finite size of the network. In finite size networks,
as $t \to \infty$ nearly all the nodes are ultimately reached and hence there
is no further 'territory' to be explored leading to deviations from Eq. \ref{ure}.
Notice also that if the agreement with scaling relation is reached faster, as
in the case of $W_1=50, W_2=100$ in Fig. \ref{dev1}, the deviations for $t >> 1$
also happen earlier in comparison with the case of, say, $W_1=20$ and $W_2=40$.
Physically, this happens because more the number of walkers, agreement with
scaling curve is reached faster and the all the nodes are visited quickly (than for
smaller number of walkers), and hence the deviation for $t>>1$ also appears quickly.

In Fig. \ref{flogw}, $s_w(t)$ is shown for a scale-free network obtained from the configuration model.
For this case too, scaled time $x(t)=Wt/N$ leads to an excellent data collapse and Eq. \ref{ure} fits
the data. The parameters $A=0.1$ and $\beta=0.2$ were estimated through regression.
The inset in Fig. \ref{flogw} shows the validity of the scaling relation in Eq. \ref{scaling}
for the random walks on configuration model for several choices of $W_1$ and $W_2$.

\begin{table}[t]
\begin{tabular}{|c|c|c|}
\toprule

Network & A & $\beta$   \\[1mm] \hline
Barabsi-Albert  & 0.724 & 0.882  \\
Model &    &        \\[1mm]
Configuration Model ($\gamma=2.2$) & 0.601 & 0.653  \\
 &    &        \\[1mm]
Enron Email Network & 0.403 & 0.652 \\[1mm]
Yeast Network & 0.564 & 0.622 \\[1mm]
Autonomous Systems & 0.601 & 0.712 \\[1mm]
Scientific Collaboration & 0.273 & 0.582 \\
Network    &      &      \\
\hline
\end{tabular}
\caption{The values of $A$ and $\beta$ in Eq. \ref{unr1} obtained through regression
for various scale-free network models.}
\label{table1}
\end{table}

\begin{figure}[t]
\subfloat[]{\includegraphics*[width=2.5in]{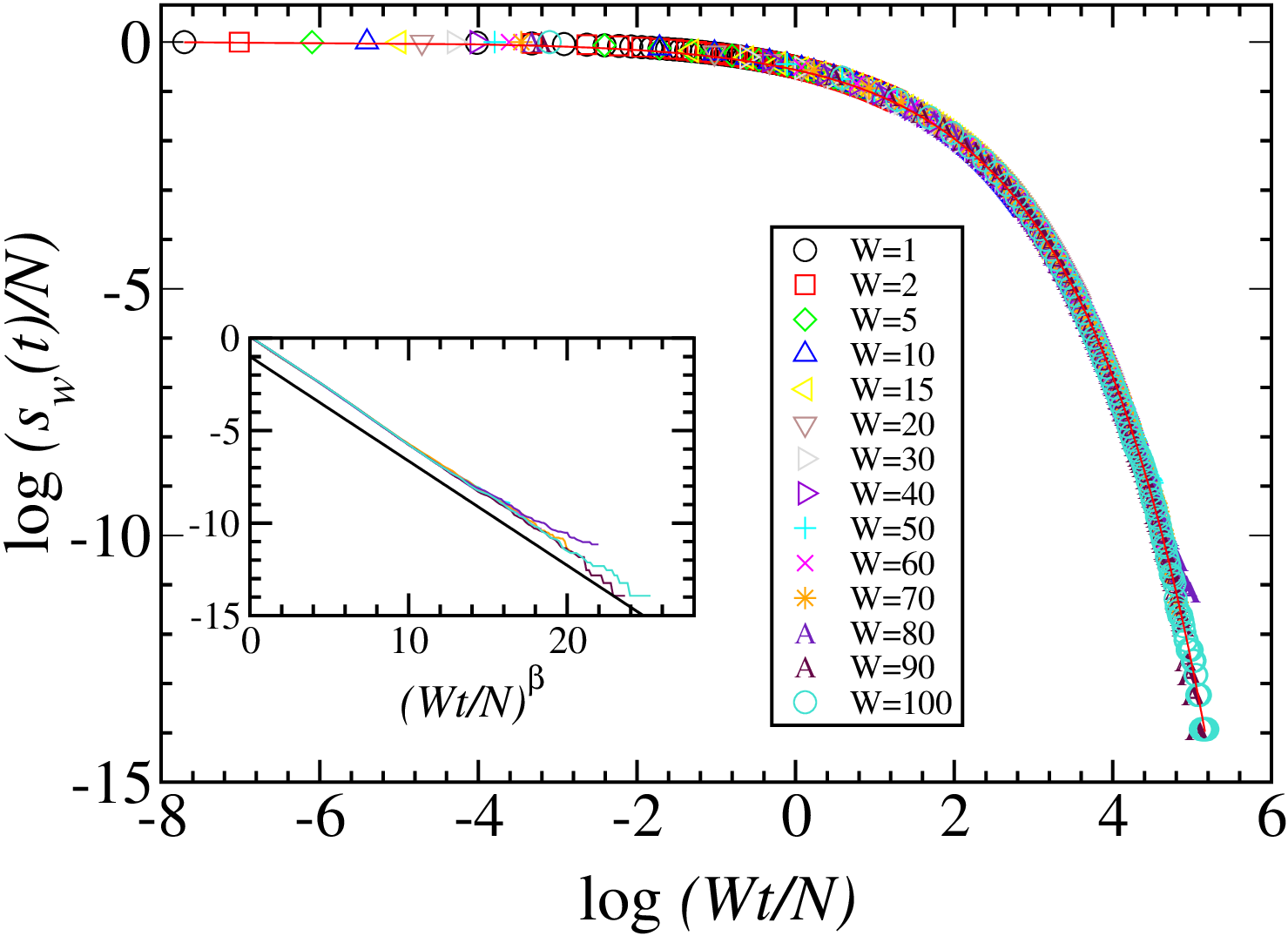}}

\subfloat[]{\includegraphics*[width=2.5in]{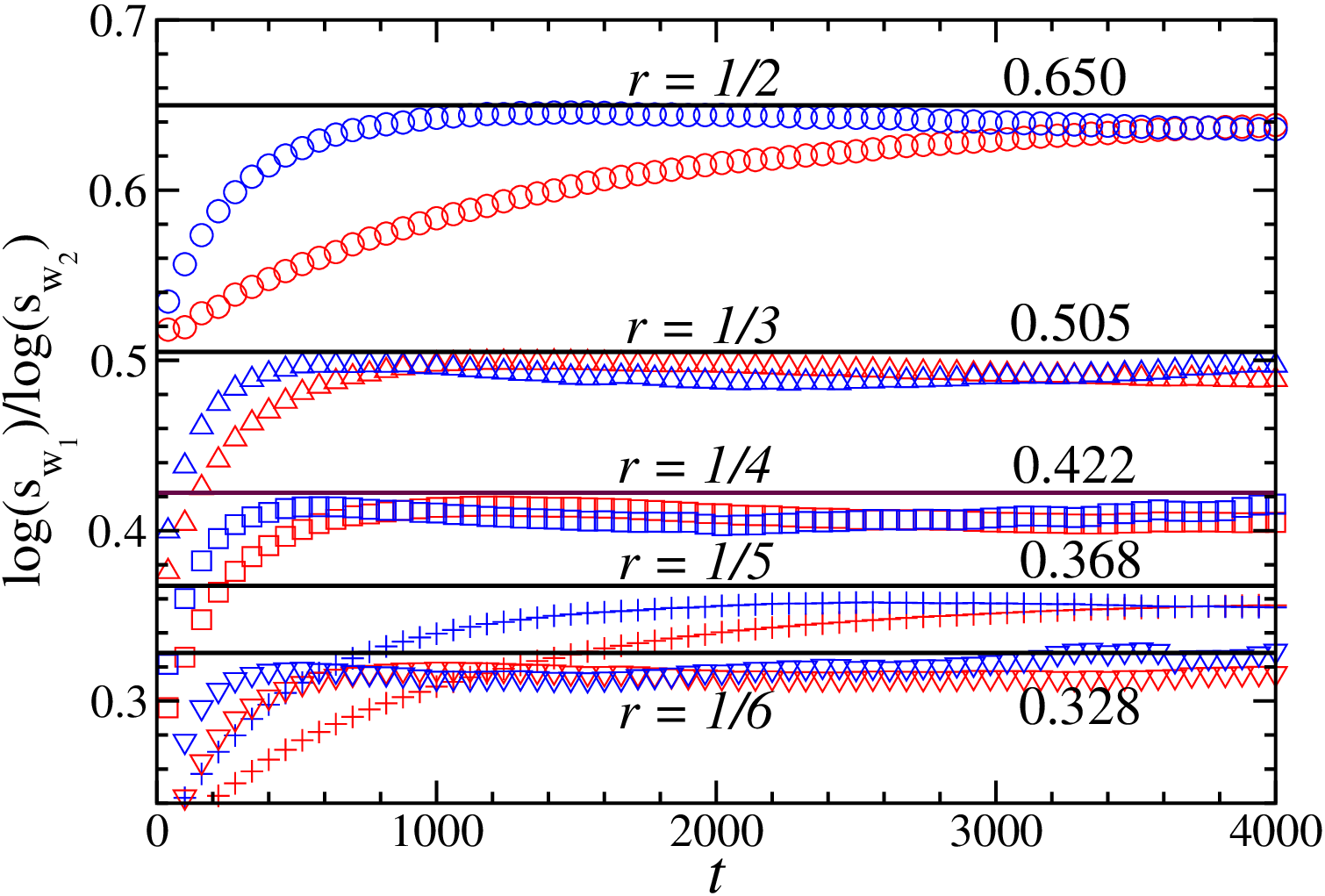}}
\caption{(a) Fraction of nodes not reached by $W$ random walkers in the network of
protein-protein interaction in yeast. The decay follows Eq. \ref{ure} with $A=0.534$
and $\beta = 0.635$. (b) The number of nodes not reached by $W_1$ and $W_2$ walkers
follows the scaling relation in  Eq. \ref{scaling}.}
\label{yeast-fig}
\end{figure}

\begin{figure}
\subfloat[]{\includegraphics*[width=2.5in]{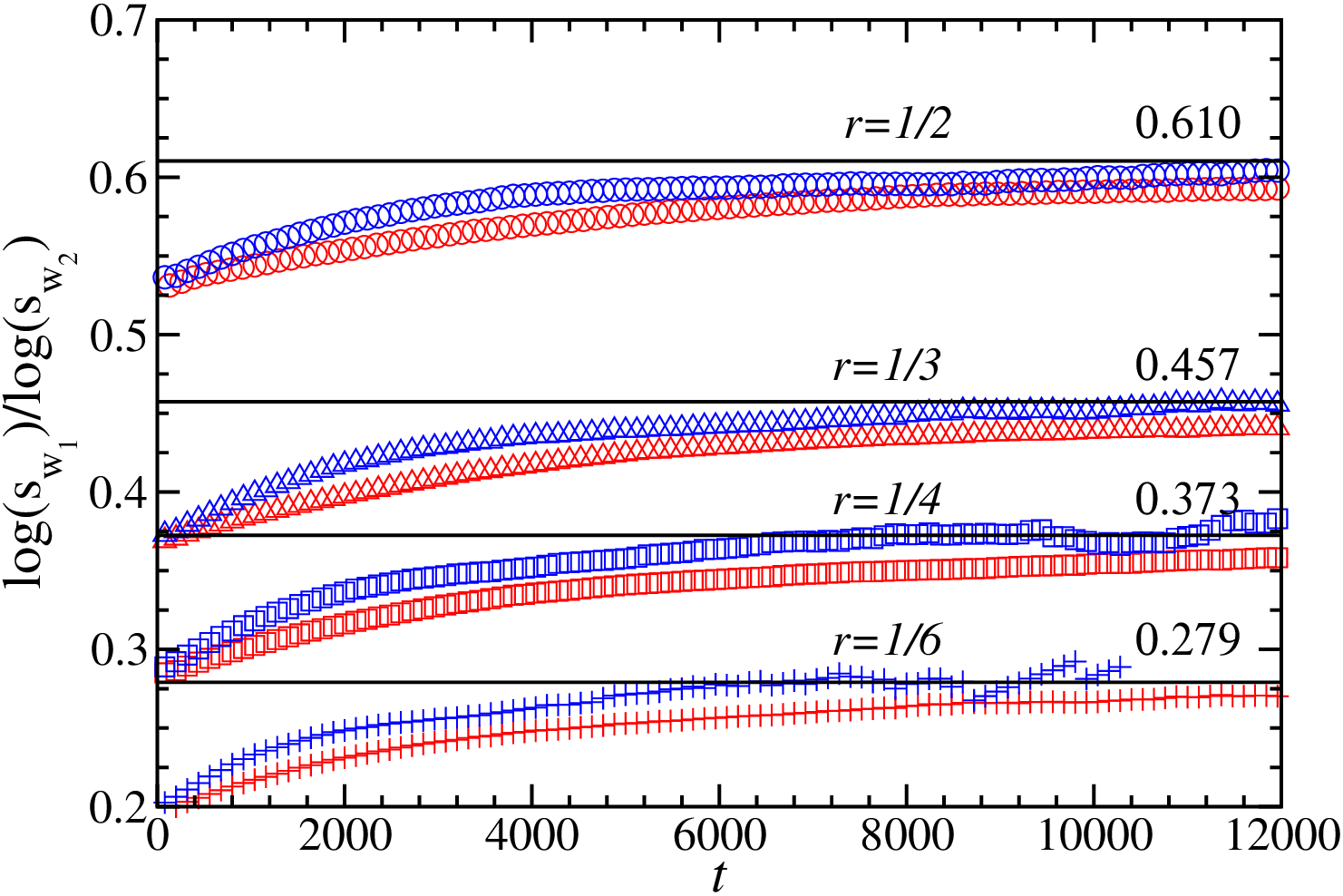}}

\subfloat[]{\includegraphics*[width=2.5in]{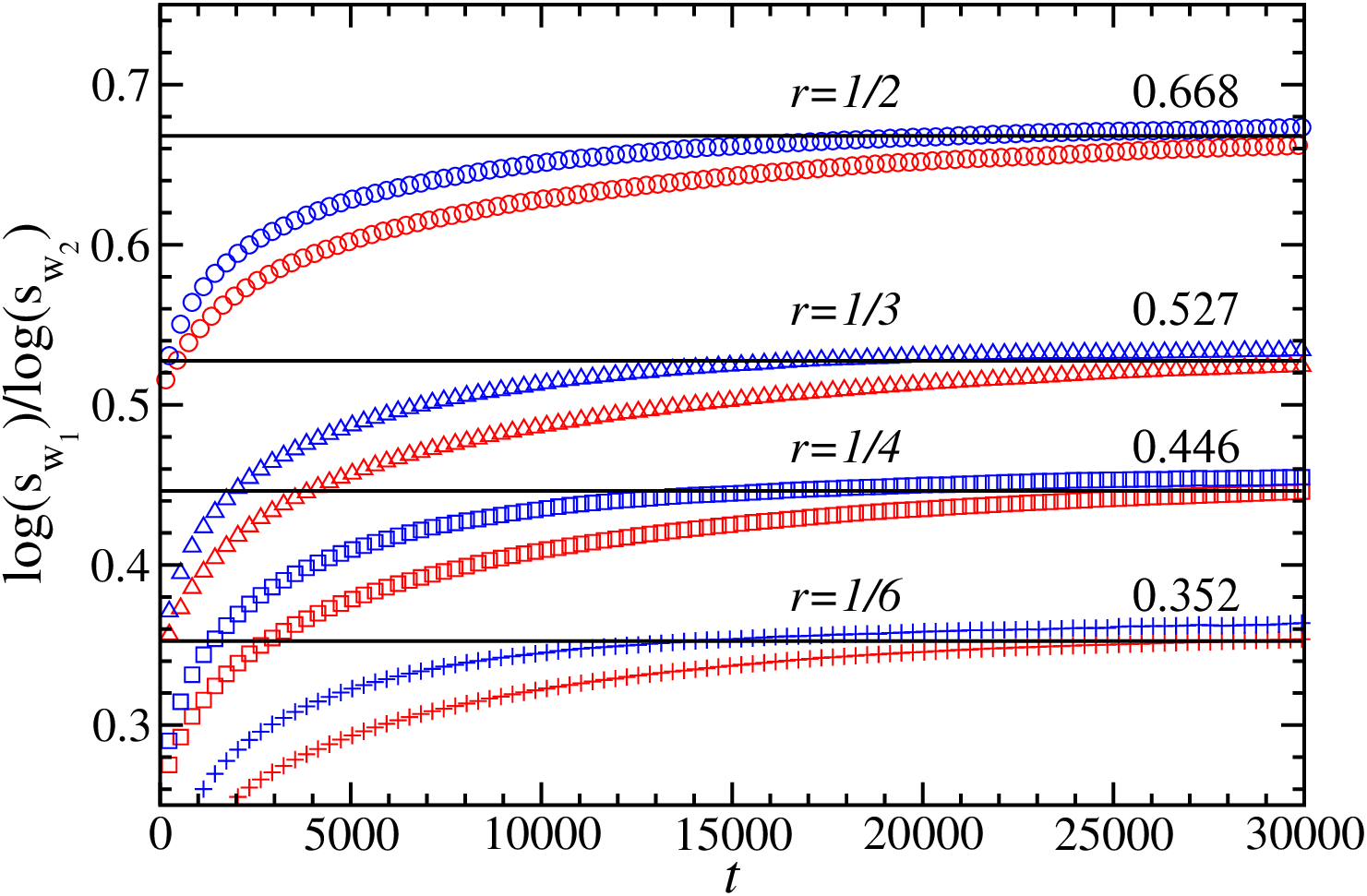}}
\caption{The ratio $u(t)$ as a function of time for random walk simulations
performed on (a) a network of autonomous systems and (b) the scientific collaboration
network of papers in cond-mat. In (a) and (b), the value of $(W_1,W_2)$ for each 
red curve is (5,10) for circles, (5,15) for up-triangles, (5,20) for squares, (5,30) 
for plus symbols. Corresponding values for blue curves are (10,20), (10,30), (10,40) 
and (10,60) respectively. The horizontal black lines correspond to $(W_1/W_2)^\beta$. 
In all the cases, numerical random walk simulations tends to this constant 
after the relaxation time $\tau_r$.}
\label{as-cond-fig}
\end{figure}

Next, we study the distinct sites visited by random walkers on four real-life networks,
namely, (a) the Enron email (EE) network and (b) protein-protein interaction network of a yeast,
(c) network of autonomous systems of the Internet connected with each other from the CAIDA 
project and (d) scientific collaboration network of cond-mat papers. We perform simulation
of random walks on these networks with $W$ walkers and the results are presented 
in Figs. \ref{enron-fig}-\ref{as-cond-fig}.
The data sets for (a,c,d) are obtained from Stanford network database \cite{snap} and
for (b) is obtained from Pajek database \cite{pajek}. All these networks were extensively studied
for their topological properties and, in particular, their degree distribution is known to
display a power-law form, $P(k) \sim k^{-\gamma}$. For Enron email network
$\gamma \approx 1.76$ \cite{koblenz}, for yeast network $\gamma \approx 2.5$ \cite{albert},
for network of autonomous systems $\gamma \approx 2.09$ \cite{koblenz} and for the network
of cond-mat papers $\gamma \approx 2.81$ \cite{koblenz}.
For the purposes of this work, the largest connected component 
of these networks was considered to ensure that isolated nodes do not exist.

Figure \ref{enron-fig}(a) shows random walk simulation results for the fraction 
of nodes not visited until time $t$ on the Enron email communication network. 
Random walk simulations were performed with different number of walkers $W$.
As this figure reveals, the simulation results are in good agreement with the 
postulated relation in Eq. \ref{ure}, with $A \approx 0.403$ and $\beta \approx 0.652$.
In this case as well, $x=Wt/N$ is the scaled time and as seen in Fig. \ref{enron-fig}(a),
an excellent data collapse is observed for number of walkers ranging from 1 to 100.
The inset in Fig. \ref{enron-fig}(a) shows the same data as a function of $(Wt/N)^\beta$
and the resulting straight line supports Eq. \ref{ure}. Further, Fig. \ref{enron-fig}(b)
shows the validity of scaling relation in Eq.\ref{scaling} for various ratio of walkers
$r = W_1/W_2$.
In Fig. \ref{yeast-fig}(a), $s_W(t)$ is shown as a function of scaled parameter for
various choices of of $W$ in log-log scale. As expected, an excellent data collapse
is observed in agreement with Eq. \ref{ure}. The inset to this figure further confirms
the temporal decay of $s_W(t)$ is indeed stretched exponential in form. As would
be expected, a good agreement with scaling relation in Eq. \ref{scaling} is seen
in Fig. \ref{yeast-fig}(b) for various ratio of walkers $r = W_1/W_2$.

The scaling results from random walker simulations on a network of autonomous
systems and author collaboration networks from cond-mat are displayed in 
Fig. \ref{as-cond-fig}. In these cases too, the simulation results for $s_w(t)$ display
an excellent data collapse when plotted as a function of $x$ (not shown here).
The values of parameters $A$ and $\beta$ estimated through regression is summarised in
Table \ref{table1} for all the networks, including the ones corresponding to Fig. \ref{as-cond-fig},
discussed in this work. A good agreement with the scaling form in Eq. \ref{scaling}
is shown in Fig. \ref{as-cond-fig}. For $t > \tau_r$, the ratio
$\log(s_{w_1})/\log(s_{w_2})$ tends to a constant dependent on the value of $\beta$,
$W_1$ and $W_2$.

In summary, we have studied the problem of distinct number of sites visited
by multiple walkers on a scale-free network. Through numerical simulations,
we have shown that the mean number of sites not reached until time $t$ 
can be represented by a stretched exponential function in Eq. \ref{ure} with $x=Wt/N$ being the
scaled parameter. Using this, we have displayed the results in the form
of a scaling relation (Eq. \ref{scaling}) between the nodes not reached in 
time $t$ by $W_1$ and $W_2$ walkers on the same network.
Thus, effectively, the problem of finding the distinct sites visited by
$W$ walkers on scale-free networks is related to that of one walker,
effectively simplifying the problem. Thus, for scale-free networks, exact results
for $s_w(t)$ for one walker would also help solve the problem for many walkers.
This results gets better as the size of the network $N$ and number of walkers $W$
tend to larger values.
We have numerically demonstrated these results for random walk dynamics on
Barabasi-Albert scale-free network and that constructed using configuration model.
Finally, we verified all our results by
simulating random walks on four different real world scale-free networks
and showing that the scaling holds. It turns out that the scale-free network 
is somewhat special as far as this scaling relations are concerned because
we did not observe such scaling relations for the other popular classes of
networks, namely, small-world and Erdos-Renyi random networks.


While the stretched exponential function is ubiquitous in the study of relaxation
processes in condensed matter systems \cite{klafter}, there are very few models in 
which stretched exponential decay of some
observable occurs naturally. We propose that this simple model of random walks on
scale-free networks with unreachability as the observable can be used as a model 
to investigate relaxation dynamics. We have seen that the stretching exponent $\beta$
varies, though not systematically, for different values of the power law exponent $\gamma$
of the degree distribution of the scale-free networks. This shows that $\beta$ has a 
dependence on the finer details of the structure of the network. 
An interesting and promising direction would be obtain analytical justifications
for the scaling relation reported in this work.



\begin{acknowledgments} 
The authors would like to acknowledge Yagyik Goswami for
the initial contribution he made to this work.
\end{acknowledgments}


\begin{thebibliography}{99}
\bibitem{rw} S. Chandrasekhar, Rev. Mod. Phys. {\bf 15}, 1 (1943);
J. Rudnik and G. Gaspari, {\it Elements of the random walk : An introduction for advanced 
students and researchers}, (Cambridge University Press, 2004); Barry D. Hughes,
{\it Random walks and random environments}, volume 1, (Clarendon Press, 1995).
\bibitem{web}  F. R. K. Chung and W. Zhao, Bolyai Soc. Math. Stud. {\bf 20}, 43 (2010).
\bibitem{rwbio} E. A. Codling, M. J. Plank, and S. Benhamou, J. R. Soc. Interface {\bf 5}, 813 (2008).
\bibitem{eco} Eugene F. Fama, American Economic Review {\bf 104}, 1467 (2014).

\bibitem{forag} G. M. Viswanathan, M. G. E. da Luz, E. P. Raposo and H. E. Stanley,
{\it The Physics of Foraging}, (Cambridge University Press, New York, 2011).
\bibitem{migr} G. M. Viswanathan, E. P. Raposo and M. G. E. da Luz,
Physics of Life Reviews {\bf 5}, 133 (2008); G. M. Viswanathan, {\it et. al.},
Nature {\bf 381}, 413 (1996).
\bibitem{inno} I. Iacopini, S. Milojevic and V. Latora, Phys. Rev. Lett. {\bf 120}, 048301 (2018).

\bibitem{insulin} S. M. Ali Tabei {\it et. al.}, PNAS {\bf 110}, 4911 (2013).
\bibitem{dna1} L. Mirny {\it et. al.}, J. Phys. A : Math. Theor. {\bf 42}, 434013 (2009);
C. Loverdo {\it et. al.}, Phys. Rev. Lett. {\bf 102}, 188101 (2009).
\bibitem{bio} R. Phillips,  J. Kondev, J. Theriot, and H. Garcia, 
{\it Physical Biology of the Cell}, (Garland Science, New York, 2013).
\bibitem{cell} T. Lu, T. Shen, C. Zong, J. Hasty and P. G. Wolynes, PNAS {\bf 103}, 16752 (2006).
\bibitem{imseg} L. Grady, IEEE Trans. on Pattern Analysis and Machine Intelligence {\bf 28}, 1768 (2006).
\bibitem{gclust} S. A. Tabrizi, A. Shakery, M. Azadpour, M. Abbasi and M. A. Tavallaie,
Physica A {\bf 392}, 5772 (2013).
\bibitem{llu} L. L\"{u} {\it et. al.}, Phys. Rep. {\bf 519}, 1 (2012).
\bibitem{mont} E. W. Montroll and G. H. Weiss, J. Math. Phys. {\bf 6}, 167 (1965); 
J. W. Haus and K. W. Kehr, Phys. Rep. {\bf 150}, 263 (1987).
\bibitem{porter} N. Masuda, M. A. Porter, R. Lambiotte, Phys. Rep {\bf 716}, 1 (2017).
\bibitem{burioni} R. Burioni and D.Cassi, J. Phys. A : Math. Gen. {\bf 38}, R45 (2005).
\bibitem{epi} M. Draief and L. Massouli,  {\it Epidemics and Rumours in Complex Networks}, 
(Cambridge University Press, New York, 2010).
\bibitem{de} A. Dvoretzky and P. Erdos, in {\it Proceedings of the Second
Berkeley Symposium on Mathematical Statistics and Probability},
(University of California, Berkeley, 1951).
\bibitem{stan} H. Larralde, P. Trunfio, S. Havlin, H. E. Stanley, and G. H.
Weiss, Phys. Rev. A {\bf 45}, 7128 (1992); {\it ibid}, Nature {\bf 355}, 423 (1992).
\bibitem{weiss} G. H. Weiss et. al., Physica A {\bf 191}, 479 (1992).
\bibitem{yuste} S. B. Yuste and L. Acedo, Phys. Rev. E {\bf 61}, 2340 (2000).
\bibitem{snm1} A. Kundu, S. N. Majumdar, and G. Schehr, Phys. Rev. Lett. {\bf 110},
220602 (2013).
\bibitem{bethe} B. D. Hughes and M. Sahimi, J. Stat. Phys. {\bf 29}, 781 (1982).
\bibitem{snm2} C. De Bacco, S. N. Majumdar, and P. Sollich, J. Phys. A {\bf 48}, 205004 (2015).
\bibitem{slee} S. Lee, Soon-Hyung Yook and Y. Kim, Physica A {\bf 387}, 3033 (2008).
\bibitem{kulk} E. Almaas, R. V. Kulkarni, and D. Stroud, Phys. Rev. E {\bf 68},
056105 (2003).
\bibitem{klafter} J. Klafter and M. F. Schlesinger, PNAS {\bf 83}, 848 (1986).
\bibitem{scale}  R. Albert and A. L. Barabási, Rev. Mod. Phys. {\bf 74}, 47 (2002).
\bibitem{newman} M. E. J. Newman, in {\it Handbook of Graphs and Networks :
From the Genome to the Internet}, (Wiley-VCH, Berlin, 2003).
\bibitem{alon} N. Alon et. al., arXiv:0705.0467
\bibitem{covtime1} B. F. Maier and D. Brockmann, Phys. Rev. E {\bf 96}, 042307 (2017).
\bibitem{snap} J. Leskovec and A. Krevl, SNAP Dataset at http://snap.stanford.edu/data (2014).
\bibitem{pajek} V. Batagelj and A. Mrvar, Pajek dataset at http://vlado.fmf.uni-lj.si/pub/networks/data/ (2006).
\bibitem{koblenz} This value provided in http://konect.uni-koblenz.de/test/networks/
\bibitem{albert} R. Albert, J. Cell Sci. {\bf 118}, 4947 (2005).

\end{thebibliography}
\end{document}